\documentclass[11pt,letterpaper]{article} 
\usepackage{hyperref}
\hypersetup{
  colorlinks   = true, 
  urlcolor     = blue, 
  linkcolor    = blue, 
  citecolor   = coolblack 
}
\pdfoutput=1
\usepackage{ol2}
\usepackage{mwe}
\usepackage{epstopdf}
\usepackage{amsmath}
\usepackage{mwe,tikz}
\usepackage[percent]{overpic}
\usepackage{verbatim} 
\usepackage{floatrow}
\usepackage[outercaption]{sidecap}
\usepackage{helvet}
\usepackage{relsize}
\usepackage{scalefnt}
\usetikzlibrary{spy}
\tikzset{fontscale/.style = {font=\relsize{#1}} }
\definecolor{coolblack}{rgb}{0.0, 0.18, 0.39}

\usepackage{fancyhdr}
\definecolor{light-gray}{gray}{0.87}

\begin{document}

\title{Beyond the current noise limit in imaging through turbulent medium}


\author{   Adam Popowicz$^1$, Aleksander Kurek$^{2,*}$, Agnieszka Pollo$^{2, 3}$, Bogdan Smolka$^1$    }

\address{
$^1$Silesian University of Technology, Institute of Automatic Control \\ Akademicka 16, 44-100 Gliwice, Poland \\
$^2$Astronomical Observatory of the Jagiellonian University, \\ Orla 171, 30-244 Krakow, Poland \\
$^3$National Centre for  Nuclear Research, \\ A. Sołtana 7, 05-400 Otwock, Poland \\
$^*$Corresponding author: aleksander.kurek@uj.edu.pl
}

\begin{abstract}
Shift-and-add is an approach employed to mitigate the phenomenon of resolution degradation in images acquired through a turbulent medium. Using this technique, a large number of consecutive short exposures is registered below the coherence time of the atmosphere or other blurring medium. The acquired images are shifted to the position of the brightest speckle and stacked together to obtain high-resolution and high signal-to-noise frame.  \\
In this paper we present a highly efficient method for determination of frames shifts, even if in a single frame the object cannot be distinguished from the background noise. The technique utilizes our custom genetic algorithm, which iteratively evolves a set of image shifts. We used the maximal energy of stacked images as an objective function for shifts estimation and validate the efficiency of the method on simulated and real images of simple and complex sources. Obtained results confirmed, that our proposed method allows for the recovery of spatial distribution of objects even only 2\% brighter than their background. The presented approach extends significantly current limits of image reconstruction with the use of shift-and-add method. The applications of our algorithm include both the optical and the infrared imaging. Our method may be also employed as a digital image stabilizer in extremely low light level conditions in professional and consumer applications.\\
\centerline{\href{ https://www.osapublishing.org/ol/abstract.cfm?uri=ol-40-10-2181 }{Original @ Optics Letters website}} \\ \\
\end{abstract}


\vspace{4mm}
\section*{Introduction}
\vspace{-1mm}
\noindent
Images obtained through a turbulent medium (like the atmosphere) suffer from serious quality degradation. There are several methods developed to alleviate this problem. One of them is the adaptive optics \cite{JB}, which is a method commonly used in astronomy, microscopy and surveillance. In this technique, a high-speed camera monitors continuously the shape of the  wave front and the closed-loop system compensates for the wave distortions by deforming a special mirror in the optical path \cite{DF}. However, this approach is very complex, expensive and requires sophisticated equipment (e.g. the deformable mirror or the wavefront sensing devices). Moreover, the required wavefront diagnostics is available only for the objects much brighter than the background. If there is not enough light to perform the wavefront analysis, only the tip-tilt correction is applied \cite{BI}. It is accomplished by oscillating an additional mirror in optical path, which results in real-time image shifting \cite{BI, SW}.

\noindent
Another, much simplified yet efficient approach, is the Lucky Imaging \cite{FL}, where consecutive short exposures are analyzed, shifted to the brightest speckle and stacked together, as depicted in Fig.~\ref{fig:f1}. Only the images with the highest Strehl ratio\footnote{The Strehl ratio is a measure of the quality of optical image formation. It has a value in the range 0 to 1, where 1 describes an unaberrated optical system.} are included, hence the final image is the average of the frames captured under best atmospheric conditions. The probability of obtaining such good quality and unblurred frames is a function of the aperture size of $D$ and the Fried parameter $r_{0}$ \cite{TI}:

\begin{figure}[!t]
\floatbox[{\capbeside\thisfloatsetup{capbesideposition={right,center},capbesidewidth=5cm}}]{figure}[\FBwidth]
{\caption{
A series of consecutive frames, which have to be shifted (red arrows) toward the common image center (green dashed line) and averaged to produce high resolution image.
}\label{fig:f1}}
{\includegraphics[width=10cm, height=4cm]{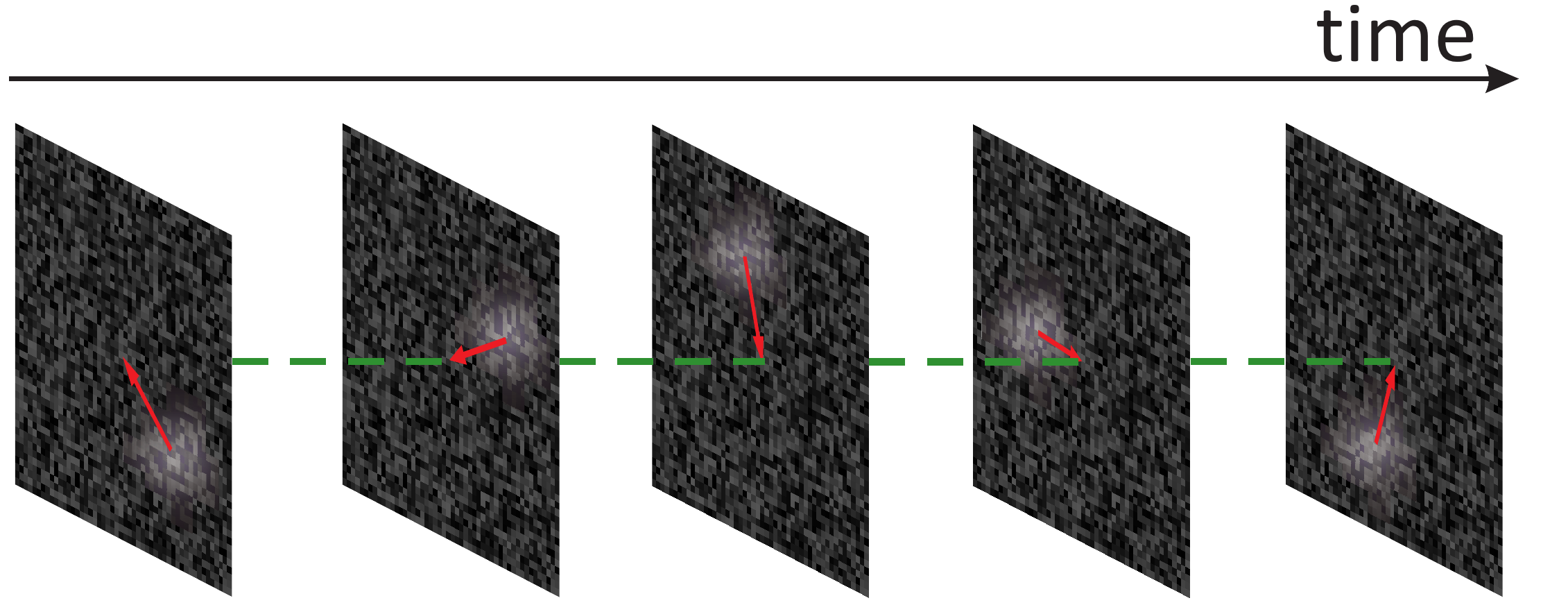}}
\end{figure}

\begin{equation}
 P \approx  5.6 \; \exp \left ( -0.1557\left ( D/r_{0} \right )^2  \right )         \quad\text{for}\;  \;  D/r_{0} \geq 3.5.
\end{equation}

\noindent
However, even those good quality frames are shifted relative to each other. The cancellation of tip-tilt component recovers about 50\% of spatial distribution in most cases \cite{CH, HC}. That kind of recovery is possible only if the source is intensive enough, so that a single frame contains clearly visible brightest speckles. Moreover, the brightest speckle varies very quickly and it is assumed to be stable for a very short, so-called coherence time (about 10\,ms) \cite{VO}. To the best of our knowledge, there were no successful attempts to recover the high quality spatial distribution of the objects using shift-and-add method, if neither the brightest speckle nor the signal centroid is visible in a single exposure.

\vspace{4mm}
\section*{Our method}
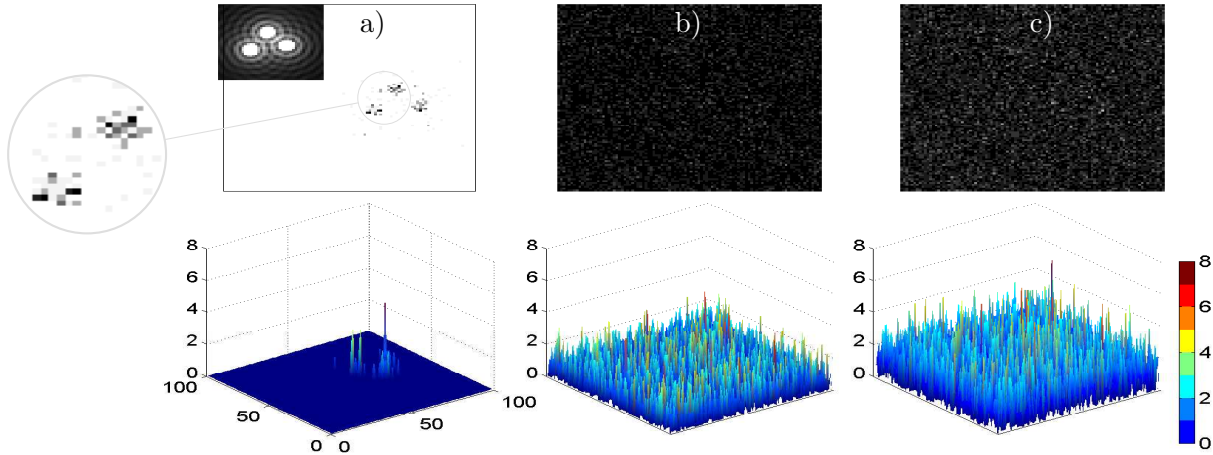
\begin{figure}
   \begin{tikzpicture}[spy using outlines={red,magnification=5,size=3cm, connect spies}]

	\node[inner sep=0pt] (russell) at (0,0)
        {\includegraphics[width=14cm, height = 6cm]{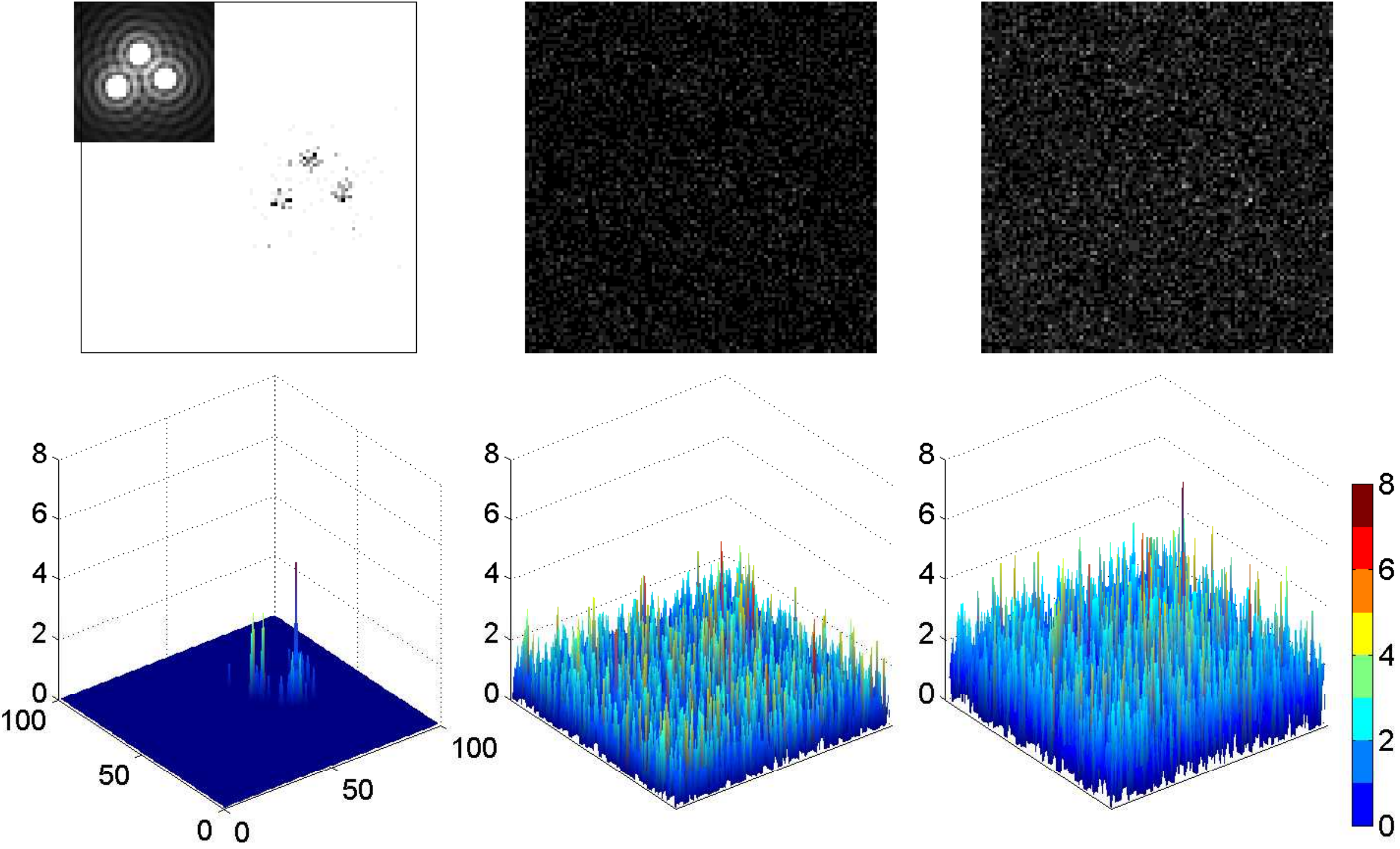}};
      \spy[circle,light-gray, thin,height = 2.1cm, width = 2.1cm, magnification = 3] on (-4.05,1.75) in node at (-8,1);
\draw (-4.2,2.4)   node [above] {a)};
\draw[white] (0,2.4)   node [above] {b)};
\draw[white] (4.7,2.4)   node [above] {c)};
    \end{tikzpicture}
\caption{
The consecutive steps of image simulation: a) -- simulated counts from the object (Airy discs model depicted in the left upper corner), b) -- object with added background counts, c) -- a final simulated image with included Gaussian electronic noise.
}
\label{fig:f2}
\end{figure}

\vspace{-1mm}
\noindent
In this paper we present a novel technique based on the genetic algorithm (GA), which allows for individual frames centring, even if the brightest speckle is not distinguishable, (see example of such frame in Fig.~\ref{fig:f2}c). Our method does not require any additional information about observed object, so it may be employed in a wide range of applications.

\noindent
GA is an iterative technique employed for the optimization purposes in highly nonlinear spaces \cite{HH,MD,DW}. It searches a solution space in order to achieve a given goal, so that in each evolution of the algorithm, a part of individuals is randomly modified in hope to produce new, better solutions. The best individuals (elite) are transferred directly between the evolutions. Another group in the population is created by the mutation of the elite. Finally, in each iteration, a part of individuals is created randomly to prevent getting stuck at a local minimum.

\noindent
The use of GA on a large number of frames allows to evolve the set of frames shifts toward the highest energy of averaged frame, i.e. -- to maximize the value of $E$ defined as:

\vspace{-2mm}

\begin{equation}
 E =   \sum_{x=1}^{N}\sum_{y=1}^{M}I_{x,y}^2,
\end{equation}

\noindent
where\, $I_{x,y}$ \,is the number of counts in a pixel at position ($x$, $y$) \ and \ \emph{N}, \emph{M} denotes the frame's width and height. We used such a fitness function, because only the perfect alignment of all images provides the maximum sum of squared counts. As the errors occur in shifts estimations, the counts from the object are spread between many pixels in the averaged stack, thus our metric decreases. Before the energy calculation, we perform background estimation utilizing a median over all pixels in stacked images. Then, the background level is subtracted, hence the influence of background counts on the energy measure $E$ is significantly reduced.

\noindent
In our simulations, first we used Airy discs as sources \cite{HF}. It is the best description of spot light that a perfect lens with circular aperture can produce. The disc size is governed by the diffraction limit, what may be observed in spaceborne telescope missions due to a lack of turbulent medium between the imager and the object (the interested reader is referred to e.g. \cite{HC2}). To simulate real count results in each pixel, we utilized Poisson distribution, where the mean value was determined from the sum of background and object counts \cite{JR1, JR2}.

\noindent
We present an exemplary result of simulated Airy discs in Fig.~\ref{fig:f2}a, where the objects produce 61 photons and the entire background counts equals 677. Finally, we combined the counts of the background with the ones from the object and added Gaussian readout noise of imager amplifier. We adjusted the level of noise and source, so that the object cannot be distinguished from the background noise (see Fig.~\ref{fig:f2}c).

\noindent
In GA we optimized two variables per each frame -- the shifts in \emph{X} and \emph{Y} directions. Such an approach produces a huge number of variables, thus a whole set of combinations cannot be tested in reasonable time even with the most powerful computers. Assuming that the object tip-tilt effect is limited to $\pm$\emph{p} pixels in each direction, we have the following number of possible shift combinations\,~$c$\, for \,$K$\, images in a series\,:

\vspace{-2 mm}
\begin{equation} \label{eq:eq3}
 c =   \frac{   \big((2p+1)^2+K-1\big)   !}  {   (2p+1)^2\, ! \ (K-1) !   },
\end{equation}

\noindent
In contrast, with the use of GA, this optimization process can be significantly reduced, since the algorithm promotes and modifies only viable shifts combinations while the others are discarded. For example, in our experiments, the recovery of 70 frames of size \,100$\times$100, took about 30 minutes (1500 generations) in the case of simple point-like sources. According to Eq. (\ref{eq:eq3}),\,\  3.75E+479 years would be required on the same computer to search a whole combinations space ($c$\,=\,1.97E+489 combinations for $K$\,=\,70 \,images in a series, maximum shift\, $p$\,=\,7; the computational time of a single shift-and-add procedure is 6\,ms).

\noindent
We used 700 individuals in our GA population. By an individual we understand the vector of shifts in $X$ and $Y$ directions for all images in a series (e.g. for 100 images the individual is 200-elements vector). In each iteration, the best solution was transferred directly from the previous population and 500 individuals were created by the crossover procedure from the shifts estimations of previous 140 individuals (elite). We employed scattered crossover, thus each shift of newly created solution was selected from a random individual from the elite. Next 100 individuals were created by mutation, where we changed the shifts of previous best individual using the Gaussian distribution with a standard deviation equal to 1/3 of maximum available shift.

\noindent
The GA final outcomes were the shifts estimations obtained after 1500 iterations. Such value was chosen due to the observed satisfactory results in this initial part of the convergence process and due to the reasonable computational time (30 minutes). It should be noted, that better outcomes may be obtained after much longer GA run. However it would require randomly varying calculation time, often reaching tens of hours.

\noindent
It is difficult to present strict numeric properties of the method, since each individual run depends on its randomly generated starting point. Nevertheless, from multiple tests we observed, that the method allows for the recovery of the spatial distribution of objects as dim, as 2\% of their mean background. For such intensity of light, about 70 exposures were required to obtain satisfactory results using our GA-based algorithm.

\begin{figure*}
\scalefont{1.5}
\begin{tikzpicture}
[       every node/.style={anchor=south west,inner sep=0pt,font=\bfseries}, x=1mm, y=1mm     ]
      \node (fig1) at (0,30)            {\includegraphics[width=17.7cm]{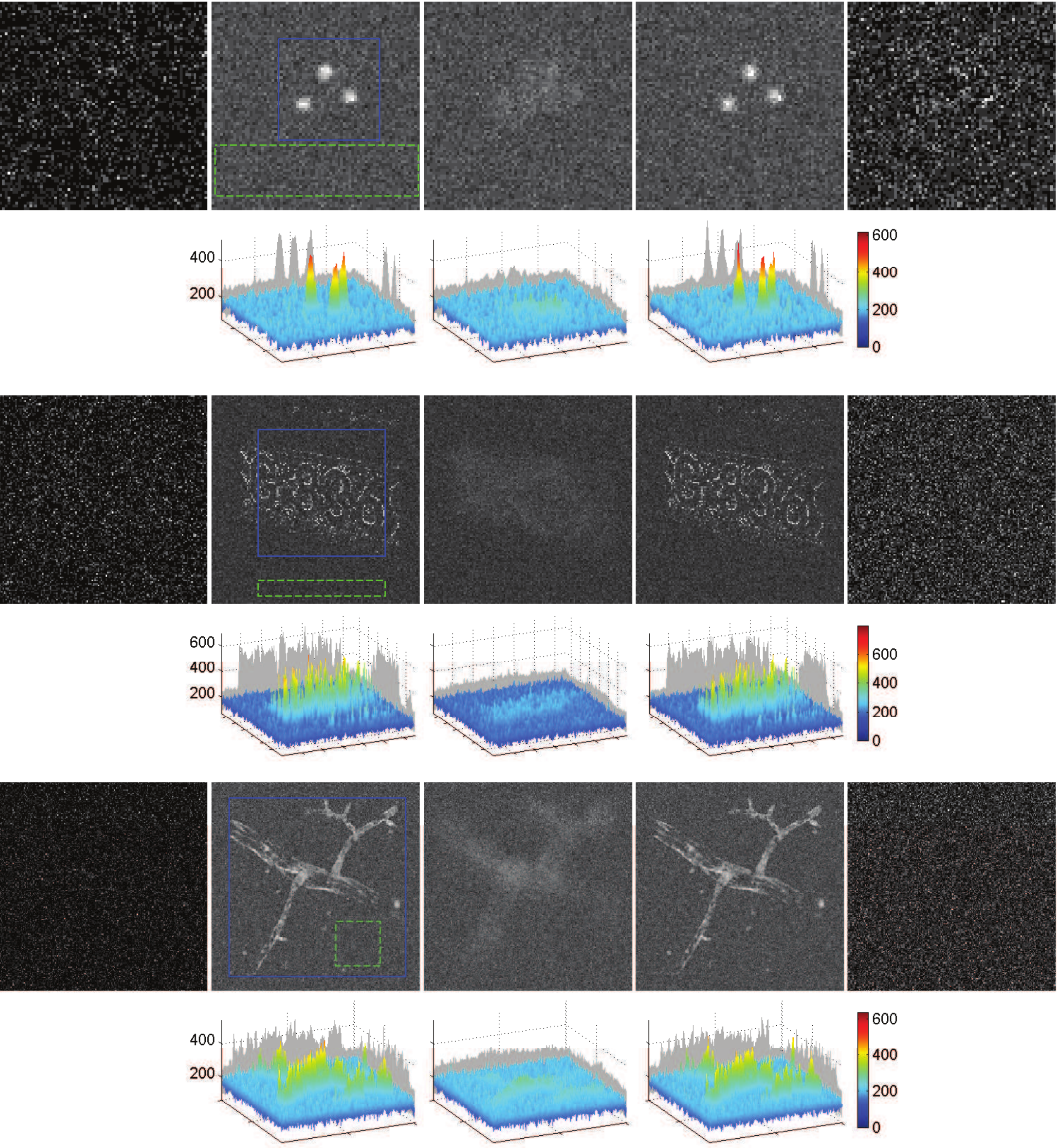}};
      5

\draw[white] (3.3,215)   node [above] {a)};
\draw[white] (39,215)   node [above] {b)};
\draw[white] (74.5,215)   node [above] {c)};
\draw[white] (110,215)   node [above] {d)};
\draw[white] (145,215)   node [above] {e)};

\draw[white] (47,192)   node [above]{\small BLUR:   0.000};
\draw[white] (83,192) node [above] {\small BLUR: 1.000};
\draw[white] (118.5,192)  node [above] {\small BLUR:   0.028};
\draw[white] (83.5,188)  node [above] {\small RMSE:   22.94};
\draw[white] (119,188)  node [above] {\small RMSE:   14.10};
\draw[white] (3,149)   node [above] {f)};
\draw[white] (39,149)  node [above] {g)};
\draw[white] (74.5,149)   node [above] {h)};
\draw[white] (110,149)  node [above] {i)};
\draw[white] (145,149)   node [above] {j)};

\draw[white](47,126)   node [above]{\small BLUR:   0.000};
\draw[white] (83,126) node [above] {\small BLUR: 1.000};
\draw[ white] (118.5,126)  node [above] {\small BLUR:   0.040};
\draw[white](83.5,122)  node [above] {\small RMSE:   22.16};
\draw[white] (119,122)   node [above] {\small RMSE:   13.53};
\draw[white] (3,85)   node [above] {k)};
\draw[white] (39,85)  node [above] {l)};
\draw[white] (75.3,85)   node [above] {m)};
\draw[white] (110,85)  node [above] {n)};
\draw[white] (145.2,85)   node [above] {o)};

\draw[white](47,61)   node [above]{\small BLUR:   0.000};
\draw[white] (83,61) node [above] {\small BLUR: 1.000};
\draw[ white] (118.5,61)  node [above] {\small BLUR:   0.041};
\draw[white](83.5,57)  node [above] {\small RMSE:   23.40};
\draw[white] (119,57)   node [above] {\small RMSE:   13.76};
\end{tikzpicture}
\caption{
The results of image enhancement employing our method:\, a), f), k) -- exemplary single frames, the contrast in these images was increased for better readability;\, b), g), l) -- average of 70 frames without simulated turbulence, green and blue rectangles represent the areas utilized for the signal-to-background measurements;\, c), h), m) -- average of 70 frames with simulated turbulence;\, d), i), n) -- average of 70 frames after applying image shifts corrections provided by our algorithm;\, e), j), o) -- the difference between the reference image and the GA enhanced result (error map).
}
\label{fig:f3}
\end{figure*}
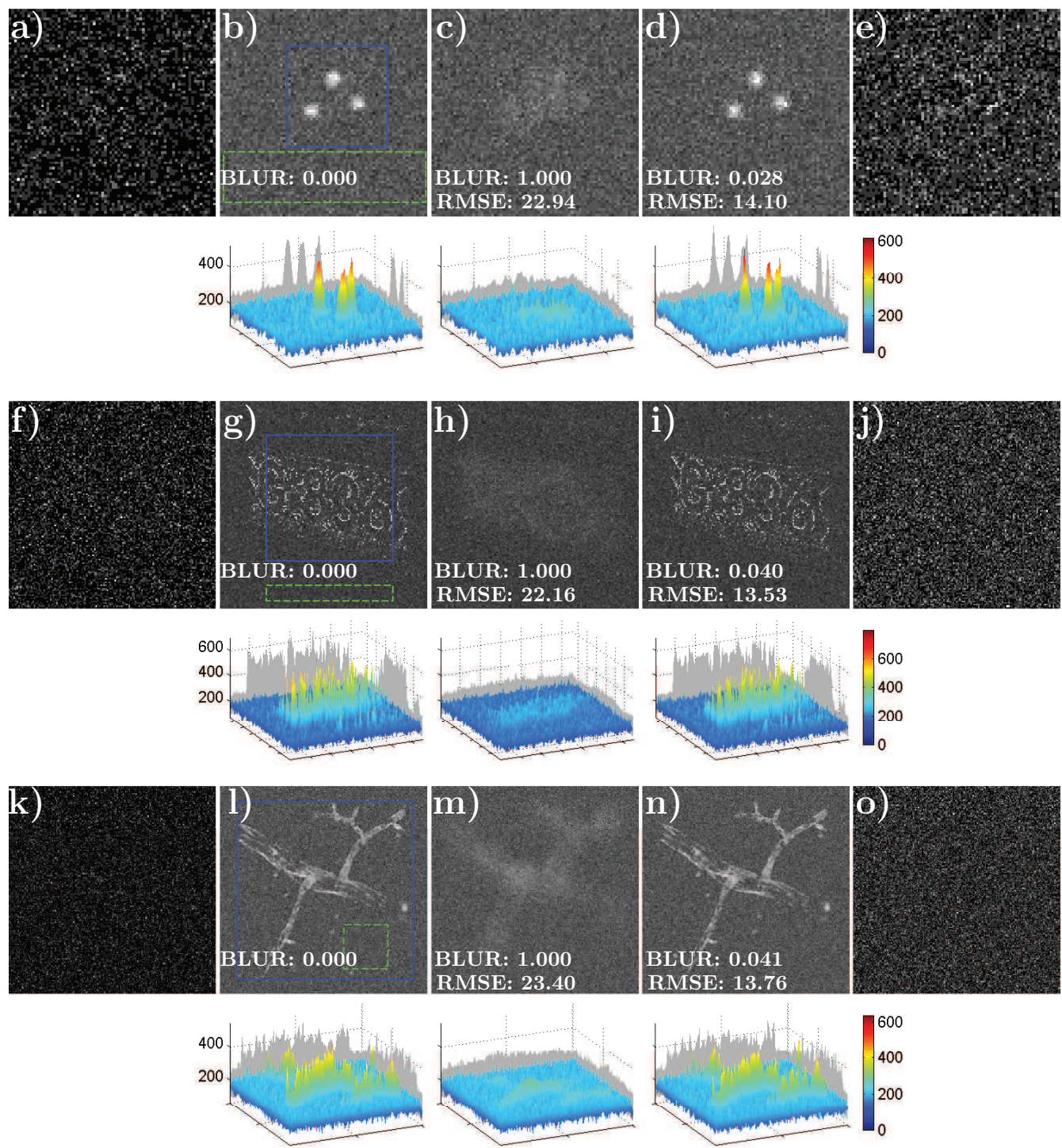

\noindent
In Fig.~\ref{fig:f3} \,we present the outcomes of our algorithm for exemplary images. The first one depicts the aforementioned point-like objects. To obtain the remaining two complex images, we cropped 100$\times$100 pixels from the image of a car licence plate and 200$\times$200 pixels of human retina. We simulated the required short exposures by reducing significantly the pixels intensities. Then, we shifted the frames randomly. We decided to select uniform distribution, so that the estimation was more challenging, than it would be for Gaussian distribution, in which a majority of small shift deviations is observed. We also accounted for the Poisson and read-out noise of imager amplifier. Then, we ran our GA procedures on such a set of frames.

\noindent
The presented experiment reflects the imaging process through the turbulent medium, where the Fried parameter $r_0$ is equal to the lens size. In such a case, the most prominent resolution degradation originates in a global tip-tilt effect within the image plane.

\noindent
In our experiments we obtained visually satisfactory results, even if the light flux was about \,2-3\% \,stronger than the background. The image enhancement is significant and the error maps contain only the noise without any visible object signature. The calculations of the signal to background level were based on the photon count, measured within green (background only) and blue (background and signal) regions (see Fig~\ref{fig:f3}b, g and l). For the point-like object, the signal-to-background ratio was 3.7\%. For the complex images it was, respectively 1.78\% and 1.92\% for the license plate and human retina.

\noindent
In Fig.~\ref{fig:f3} we also included the results of frequently used image quality measures: RMSE (Root-Mean-Square Error, in counts) and BLUR (see \cite{CD} for description). For better readability, BLUR outcome was normalized, so that for the reference its value is  0, and for the stack with simulated turbulence, it is 1. To reduce the impact of background noise on results of quality measures, all images were smoothed using a\, 3$\times$3 \,moving average. The reference for the comparisons was obtained by stacking images without simulated turbulence.


\begin{figure}[!t]
\floatbox[{\capbeside\thisfloatsetup{capbesideposition={right,center},capbesidewidth=4cm}}]{figure}[\FBwidth]
{\caption{
Shift canceling displacement computed by GA vs. real displacement -- the results of an exemplary experiment run performed on 70 frames of point-like objects depicted in Fig.~\ref{fig:f2}.
}\label{fig:tf4}}
{\includegraphics[width=12cm]{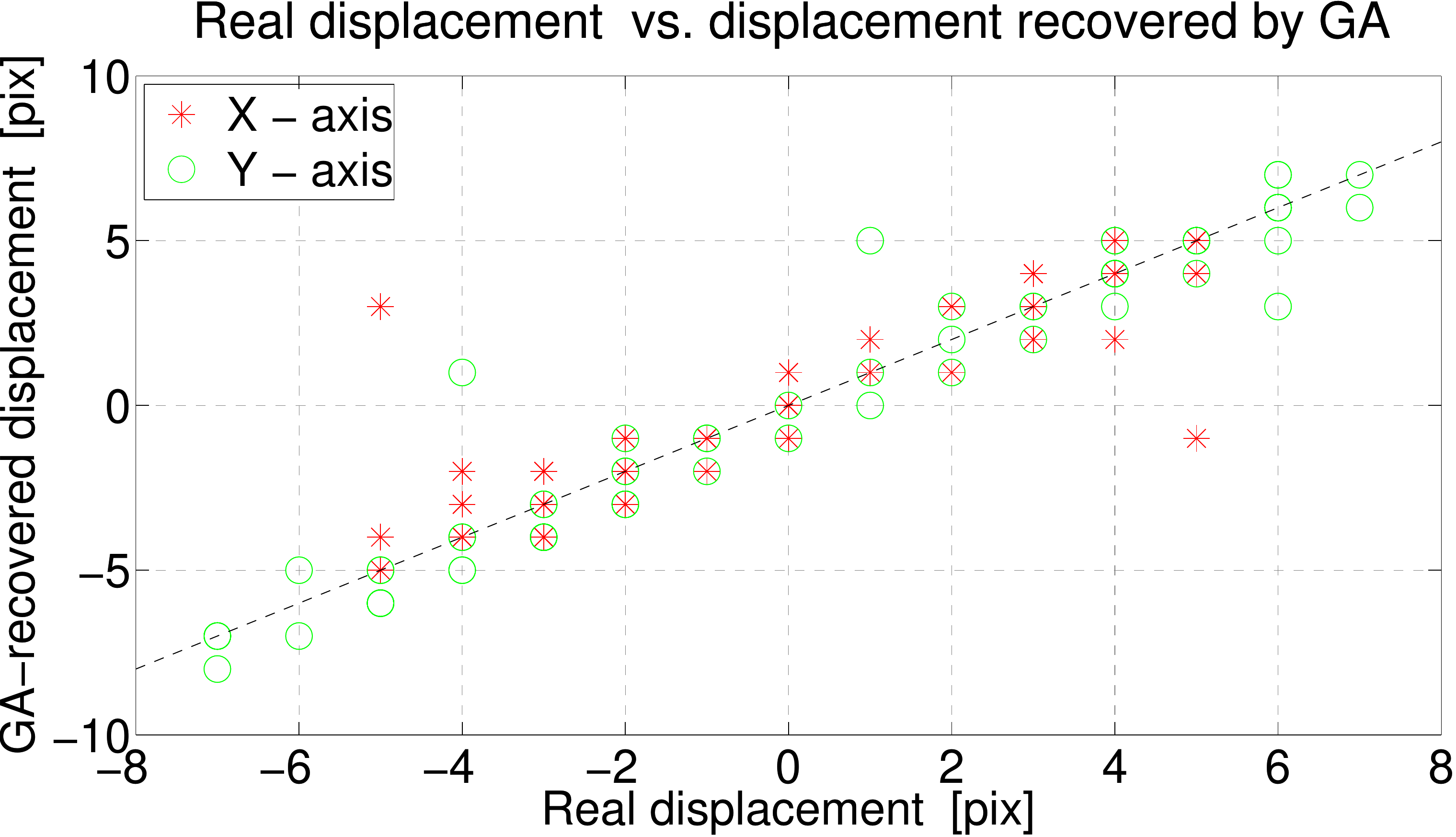}}
\end{figure}

\noindent
In Fig.~\ref{fig:f4} \ we present a comparison between calculated and simulated \ \emph{X} \ and \ \emph{Y} \ shifts in an exemplary experiment run. The calculated mean shift error was, respectively, \ 0.13 \ and \ -0.2 \ for \emph{X} and \emph{Y} direction. The standard deviation was \ 1.43 (\emph{X}) and \ 1.08 (\emph{Y}). As predicted, for the most of the frames, the shifts estimations provided by our method were the same or very close to the reference ones.

\vspace{4 mm}
\section*{Summary}
\noindent
In this paper we demonstrated the first results of spatial distribution recovery in shift-and-add method when no brightest speckle nor the signal centroid is visible in a frame. The spatial distribution, to the best of our knowledge, not possible to resolve by any other approach, was successfully reconstructed. The proposed GA method may be employed in a wide range of applications including astronomy, medicine or surveillance. Moreover, it is an automatic method and no knowledge about spatial distribution, number of sources or their position is required. It is also very easy to implement, as GA solvers are presently available in numerous software packages.

\noindent
The proposed technique is suitable for real and simulated data in both optical and infrared bands. Our method may be also employed as a digital image stabilizer in extremely low light level conditions in both professional and consumer applications.

\vspace{4 mm}
\subsection*{Acknowledgments}
\noindent
Adam Popowicz was supported by Polish National Science Center, grant no. 2013/11/N/ST6/03051: Novel Methods of Impulsive Noise Reduction in Astronomical Images. The research was performed using the infrastructure supported by POIG.02.03.01-24-099/13 grant: GeCONiI -- Upper Silesian Center for Scientific Computation. This research was supported by the Dean of the Faculty of Physics, Astronomy and Applied Computer Sciences of the Jagiellonian University, decision no. 7150/E-338/M/2014. This research was also supported by the National Science Center grant UMO-2012/07/B/ST9/04425.

\vspace{6 mm}

\begingroup
\renewcommand{\section}[2]{}%

\endgroup


\subsection*{\textbf{References}}
\vspace{1 mm}
\begin{thebibliography}{99}

\bibitem{JB} J. M. Backers, ``Adaptive optics for astronomy -- Principles, performance, and applications'',  Annu. Rev. Astron. Astr. \textbf{31,} 13-62 (1993)
\href{  http://adsabs.harvard.edu/abs/1993ARA%26A..31...13B  }{online version}

\bibitem{DF} R. W. Duffner, R. Q. Fugate, ``The Adaptive Optics Revolution: A History'', University of New Mexico Press (2009)
\href{  http://www.unmpress.com/books.php?ID=12130420126788  }{online version}

\bibitem{BI} N. Baba, S. Isobe, et al., ``Stellar speckle image reconstruction by the shift-and-add method", Appl. Optics.. \textbf{24,} 1403-1405 (1985)
\href{  http://ukads.nottingham.ac.uk/cgi-bin/nph-bib_query?bibcode=1985ApOpt..24.1403B&db_key=AST  }{online version}

\bibitem{SW} A. Sivaramakrishnan, R. Weymann, et al., ``Measurements of the Angular Correlation of Stellar Centroid Motion", Astrophys. J. \textbf{110,} 430-438 (1995)
\href{  http://adsabs.harvard.edu/abs/1995AJ....110..430S  }{online version}

\bibitem{FL} D. L. Fried, ``Probability of getting a lucky short-exposure image through turbulence", J. Opt. Soc. Am. \textbf{68,} 1651-1658 (1978)
\href{  http://adsabs.harvard.edu/abs/1978OSAJ...68.1651F  }{online version}

\bibitem{TI} V. Tikhomirov, ``Dissipation of energy in isotropic turbulence", Proc. R. Soc. A \textbf{25,} 324-327 (1991)
\href{  http://link.springer.com/chapter/10.1007%2F978-94-011-3030-1_47  }{online version}

\bibitem{CH} J. Christou, ``Image quality, tip-tilt correction, and shift-and-add infrared imaging", Publ. Astron. Soc. Pac. \textbf{103,} 1040-1048 (1991)
\href{  ttp://adsabs.harvard.edu/abs/1991PASP..103.1040C  }{online version}

\bibitem{HC} J. Hecquet, G. Coupinot, ``A gain in resolution by the superposition of selected recentered short exposures", J. Opt. \textbf{16,} 21-26 (1985)
\href{  http://adsabs.harvard.edu/abs/1985JOpt...16...21H  }{online version}

\bibitem{VO} V. Voitsekhovich, V. Orlov, ``Temporal properties of the brightest speckle", Rev. Mex. Astron. Astr. \textbf{50,} 37-40 (2014)
\href{  http://www.redalyc.org/articulo.oa?id=57131044005  }{online version}

\bibitem{HH} R. Haupt, S. Haupt, ``Practical Genetic Algorithms'', John Wiley \& Sons (1998)
\href{  http://library.alibris.com/Practical-Genetic-Algorithms-Randy-L-Haupt-PH-D/book/5278379?qsort=c&matches=33  }{online version}

\bibitem{MD} M. D. Vose, ``The Simple Genetic Algorithm: Foundations and Theory", MIT Press (1998)
\href{  http://dl.acm.org/citation.cfm?id=521897  }{online version}

\bibitem{DW} D. Whitley, ``A genetic algorithm tutorial", Stat. Comput. \textbf{4,} 65-85 (1994)
\href{  http://link.springer.com/article/10.1007%2FBF00175354  }{online version}

\bibitem{HF} J. Harvey, C. Ftaclas, ``Diffraction effects of telescope secondary mirror spiders on various image-quality criteria", Appl. Optics. \textbf{34,} 6337 (1995)
\href{  http://adsabs.harvard.edu/abs/1995ApOpt..34.6337H  }{online version}

\bibitem{HC2} J. Holberg, S. Casewell, ``Hubble Space Telescope imaging and spectroscopy of the sirius-like triple star system HD 217411", Mon. Not. R. Astron. Soc. \textbf{444,} 2022-2030 (2014)
\href{  http://adsabs.harvard.edu/abs/2014MNRAS.444.2022H  }{online version}

\bibitem{JR1} J. Janesick, ``Photon Transfer", SPIE Press Monograph Vol. PM170 (2007)
\href{  http://www.amazon.com/Photon-Transfer-Press-Monograph-PM170/dp/0819467227  }{online version}

\bibitem{JR2} J. Janesick, ``Scientific Charge-Coupled Devices", SPIE Press Monograph Vol. PM83 (2001)
\href{  http://www.amazon.com/Scientific-Charge-Coupled-Devices-Press-Monograph/dp/0819436984  }{online version}

\bibitem{CD} F. Crete, T. Dolmiere, et al., ``The blur effect: perception and estimation with a new no-reference perceptual blur metric", Soc. Photo.-Opt. Instru. \textbf{6492,} EI 6492-16 (2007)
\href{  http://proceedings.spiedigitallibrary.org/proceeding.aspx?articleid=1298489  }{online version}

\end{thebibliography}
\end{document}